\documentclass[aps,superscriptaddress,nofootinbib,12pt]{revtex4-2}
\usepackage[utf8]{inputenc}
\usepackage[colorlinks=true,
breaklinks=true,
urlcolor=magenta,
citecolor=blue]{hyperref}
\usepackage{graphicx}
\usepackage{bm,amsmath,amssymb}
\usepackage{color}
\usepackage{xcolor}
\newcommand{\ee}{e^+e^-}
\newcommand{\jppb}{J/\psi p\bar p}
\newcommand{\amp}{\mathcal{A}}

\newcommand{\itp}{\affiliation{Institute of Theoretical Physics, Chinese Academy of Sciences, Beijing 100190, China}}
\newcommand{\ucas}{\affiliation{School of Physical Sciences, University of Chinese Academy of Sciences, Beijing 100049, China}}

\newcommand{\peng}{\affiliation{Peng Huanwu Collaborative Center for Research and Education,\\ Beihang University, Beijing 100191, China}}

\begin{document}

\title{Production of $J/\psi p \bar p$ in electron-positron collisions}
\author{Xiao-Yu Zhang}\itp\ucas
\author{Yi-Lin Song}\email{songyilin@itp.ac.cn}\itp \ucas 
\author{Feng-Kun Guo}\email{fkguo@itp.ac.cn}\itp \ucas\peng

\begin{abstract}
    The discoveries of the hidden-charm $P_c$ pentaquarks by the LHCb Collaboration have been not confirmed by other experiments yet. 
    It is desirable to investigate the feasibility of observing them in other experiments.
    We present an order-of-magnitude estimate of the cross section of the $e^+e^-\to J/\psi p \bar p$ process, where the hidden-charm pentaquarks may be searched for, and find it to be of $\mathcal{O}(4~\text{fb})$ when the $e^+e^-$ c.m. energy is about between 6 and 7~GeV. The cross section for $e^+e^-\to P_c \bar p$ is then estimated to be $\lesssim \mathcal{O}(0.1~\text{pb})$. Thus, it is promising to search for $P_c$ at the future high-luminosity super tau-charm facilities with annual integrated luminosity about 1~ab$^{-1}$. 
    We also predict the branching fraction of $\Upsilon\to J/\psi p\bar p$ to be of $\mathcal{O}(2\times 10^{-6})$, which can be tested at $B$ factories.
\end{abstract}

\maketitle

\bigskip

In 2015, the LHCb Collaboration discovered a narrow structure, $P_c(4450)$, in the $J/\psi p$ invariant mass distribution of the $\Lambda_b\to K J/\psi p$ decay~\cite{LHCb:2015yax}. With higher statistics, the $P_c(4450)$ was found to consist of two narrower structures, $P_c(4440)$ and $P_c(4457)$, and another narrow structure, $P_c(4312)$, was observed~\cite{LHCb:2019kea}. Being excellent candidates for hidden-charm pentaquarks, they have attracted much attention; for reviews, see~\cite{Esposito:2016noz,Lebed:2016hpi,Guo:2017jvc,Olsen:2017bmm,Karliner:2017qhf,Liu:2019zoy,Brambilla:2019esw,Guo:2019twa,Chen:2022asf,Meng:2022ozq, Liu:2024uxn, Chen:2024eaq, Wang:2025sic, Doring:2025sgb}. 
However, so far no independent experimental confirmations have been made. 
GlueX searched for them in the $J/\psi$ near-threshold photoproduction process and found no evidence~\cite{GlueX:2019mkq}.
The Belle Collaboration also found no evidence of the $P_c$ states in the $J/\psi p$ invariant mass distribution of the inclusive $\Upsilon(1S,2S)$ decays~\cite{Belle:2024mcb}.

To obtain independent confirmation and understand the nature of the $P_c$ states, it is desirable to estimate the reaction rates of processes that can be used to search for them. 
The maximum energy reach of BEPCII has recently been upgraded to 5.6~GeV (BEPCII-U)~\cite{BESIII:2022mxl}, and there are proposals of super tau-charm facilities with $e^+e^-$ center-of-mass (c.m.) energies up to 7~GeV~\cite{Charm-TauFactory:2013cnj,Achasov:2023gey}. With such energies, it is possible to produce $e^+e^-\to J/\psi p \bar p$, $\Lambda_c \bar D^{(*)} \bar p$, $\Sigma_c^{(*)} \bar D^{(*)} \bar p$, and so on, and thus to search for hidden-charm pentaquarks in $e^+e^-$ collisions. Here we estimate the cross section of $e^+ e^- \to J/\psi p \bar p$ at the order-of-magnitude level.

\begin{figure}[tb]
  \centering
  \includegraphics[width=0.67\textwidth]{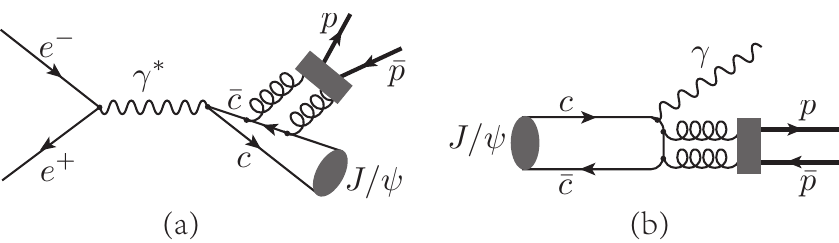}
  \caption{Two-gluon exchange mechanisms for the $e^+e^-\to J/\psi p \bar p$ and $J/\psi \to \gamma p \bar p$. The recangles stand for the hadronization of the gluons into the proton-antiproton pair.}
  \label{fig:diagrams}
\end{figure}

Because the charm quark is much heavier than the up and down quarks, the dominant mechanism for producing the $\jppb$ in electron-positron collisions should be through the electromagnetic current $\bar c\gamma_\mu c$ (see, e.g., Refs.~\cite{Keung:1980ev,Cho:1996cg, Li:2013yka}).\footnote{{The amplitude for the mechanism in which light quark-antiquark pairs are first produced through the virtual photon and $c\bar c$ is subsequently produced via gluon emission is relatively suppressed by $\alpha_s(Q^2=(2m_c)^2)$.}} 
Then the proton-antiproton pair can be produced by the gluons emitted from the charm and/or anticharm quarks.
This mechanism is shown in Fig.~\ref{fig:diagrams}~(a).
Schematically, the amplitude for the mechanism may be written as
\begin{align}
    \amp(e^+e^-\to J/\psi p\bar p) \sim \amp(e^+e^-\to J/\psi gg)\, \otimes \, \amp(gg\to p\bar p) \,, 
\end{align}
where on the right side we spell out only the amplitudes of the subprocesses $e^+e^-\to J/\psi gg$ and $gg\to p\bar p$ and omit the gluon Green's functions which are not essential in the arguments.
The first part can be computed using nonrelativistic quantum chromodynamics (NRQCD)~\cite{Bodwin:1994jh}.
The hadronization of two gluons into the proton-antiproton pair also plays a role in the radiative decay of a heavy quarkonium into $\gamma p \bar p$. 
For the $J/\psi \to \gamma p\bar p$, we have 
\begin{align}
    \amp(J/\psi \to \gamma p\bar p) \sim \amp(J/\psi \to \gamma gg)\, \otimes \, \amp(gg\to p\bar p) \,,
\end{align}
with the mechanism shown in Fig.~\ref{fig:diagrams}~(b).
Combining the above two expressions, we obtain the following estimate,
\begin{align}
  \frac{\sigma(e^+e^-\to J/\psi p\bar p)}{\Gamma(J/\psi \to \gamma p\bar p)} \approx \frac{\sigma(e^+e^-\to J/\psi gg)}{\Gamma(J/\psi \to \gamma gg)} ,
  \label{eq:ratio1}
\end{align}
where we have neglected the difference in the phase spaces, which can introduce a sizeable correction in the $\ee$ c.m. energy region just above the $\jppb$ threshold but {is weaker for energies well above it. In fact, the ratio of phase space volumes (PSVs) for the four involved processes 
$$
\frac{\text{PSV}(e^+e^- \to J/\psi p\bar p) / \text{PSV}(J/\psi \to \gamma p\bar p)}{\text{PSV}(e^+e^- \to J/\psi gg) / \text{PSV}(J/\psi \to \gamma gg)}
$$
ranges from 1.1 to 2.1 for $e^+e^-$ c.m. energies between 6 and 7 GeV.
This justifies the use of Eq.~\eqref{eq:ratio1} for an order-of-magnitude estimate in that energy region.
} 

Using the experimentally measured branching fractions~\cite{Eaton:1983kb, CLEO:2008gct, ParticleDataGroup:2024cfk},
\begin{align}
  \mathcal{B}(J/\psi \to \gamma gg) &= (8.8\pm1.1)\%\,,\nonumber\\
    \mathcal{B}(J/\psi \to \gamma p\bar p) &= (3.8\pm1.0)\times10^{-4}\,,
    \label{eq:br_jpsi}
\end{align}
we get 
\begin{align}
  \sigma(e^+e^-\to J/\psi p\bar p) \approx \sigma(e^+e^-\to J/\psi gg)\times 4\times 10^{-3}.
\end{align}

\begin{figure}[tb]
    \centering
    \includegraphics[width=0.6\linewidth]{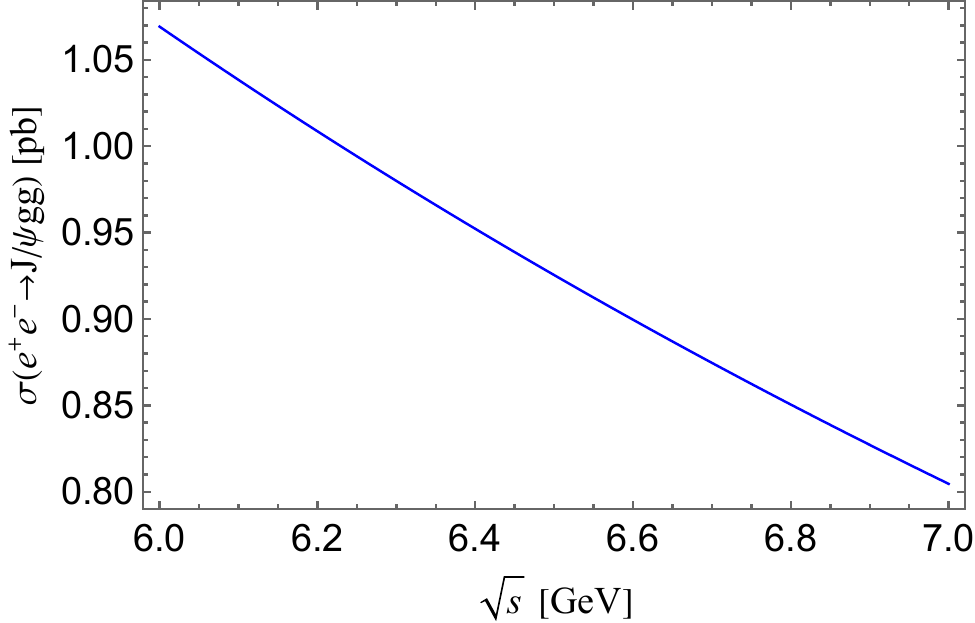}
    \caption{Cross section estimate of $\ee\to J/\psi gg$ using the LO NRQCD.}
    \label{fig:LOJpsigg}
\end{figure}
The cross section $\sigma(e^+e^-\to J/\psi gg)$ has been extensively studied in the framework of NRQCD; see, e.g., Refs.~\cite{Cho:1996cg,Yuan:1996ep,Yuan:1997sn,Baek:1998yf,Hagiwara:2004pf,Ma:2008gq}. 
The result is about 0.5~pb at the $\ee$ c.m. energy $\sqrt{s}=10.6$~GeV using NRQCD up to the next-to-leading order~\cite{Ma:2008gq}. 
Since we aim at an order-of-magnitude estimate between about 6 and 7 GeV, we use leading order (LO) NRQCD framework (for the details, see Ref.~\cite{Yuan:1997sn} and Appendix.~\ref{app:A}).
The results are shown in Fig.~\ref{fig:LOJpsigg}, with charm quark mass chosen as $m_c = 1.5$~GeV, the strong interaction coupling $\alpha_s(2m_c) = 0.26$ and the color-singlet matrix element $\langle \mathcal{O}_1^\psi(^3S_1)\rangle = 1.45 \text{ GeV}^3$~\cite{Ma:2008gq}.
Thus, for the $\ee$ c.m. energy between about 6 and 7 GeV, one has the order-of-magnitude estimate,
\begin{align}
 \sigma(e^+e^-\to J/\psi p\bar p) =\mathcal{O}(4~\text{fb}).
  \label{eq:prediction}
\end{align}

We can also predict the branching fraction of the $\Upsilon\to \jppb$, which should dominantly proceed through annihilating the $b\bar b$ into a virtual photon and is similar to the mechanism shown in Fig.~\ref{fig:diagrams}~(a).
Analogously to Eq.~\eqref{eq:ratio1}, one has
\begin{align}
  \frac{\Gamma(\Upsilon\to J/\psi p\bar p)}{\Gamma(J/\psi \to \gamma p\bar p)} \approx \frac{\Gamma(\Upsilon\to J/\psi gg)}{\Gamma(J/\psi \to \gamma gg)} .
\end{align}
Using the branching fractions in Eq.~\eqref{eq:br_jpsi} and approximating that of $\Upsilon\to J/\psi gg$ by~\cite{ParticleDataGroup:2024cfk} 
\begin{align}
    \mathcal{B}(\Upsilon\to J/\psi +\text{anything}) &= (5.4\pm0.4)\times10^{-4} \,,
\end{align}
we get
\begin{align}
    \mathcal{B}(\Upsilon\to J/\psi p\bar p) &\approx
    \mathcal{B}(\Upsilon\to J/\psi gg) \times
    \frac{\mathcal{B}(J/\psi \to \gamma p\bar p)}{\mathcal{B}(J/\psi \to \gamma gg)} \nonumber\\
    &= \mathcal{O}(2\times 10^{-6})\,.
\end{align}
{This branching fraction has not yet been measured, and} the prediction can be checked at Belle (II).

With an annually integrated luminosity of 1~ab$^{-1}$ at the Super Tau-Charm Facility (STCF) which has a designed peaking luminosity of $0.5\times10^{35} \mathrm{~cm}^{-2}\mathrm{s}^{-1}$ or higher~\cite{Achasov:2023gey} (the Super Charm-Tau Factory being developed in Novosibirsk has a similar designed luminosity~\cite{Charm-TauFactory:2013cnj}), the estimate in Eq.~\eqref{eq:prediction} leads to an expectation of $\mathcal{O}(4\times 10^3)$ $\jppb$ events per year. With such an event sample, this process can be exploited for the search of hidden-charm pentaquarks.  Assuming that the branching fraction of the $P_c\to J/\psi p$ to be at the percent level, we expect 
\begin{align}
  \sigma(e^+e^-\to P_c \bar p) \lesssim \frac{\sigma(e^+e^-\to J/\psi p\bar p)}{\mathcal{B}(P_c\to J/\psi p)} = \mathcal{O}(0.1~\text{pb}),
\end{align}
and correspondingly $\lesssim \mathcal{O}(10^5)$ $P_c\bar p$ events per year.

The $P_c$ states should also be searched for in reactions with open-charm final states, such as $e^+e^-\to  \Lambda_c \bar D^{(*)} \bar p$ and  $\Sigma_c^{(*)} \bar D^{(*)} \bar p$.  There are three reasons.
First, the hidden-charm pentaquarks are expected to decay much more easily into open-charm final states such as $\Lambda_c\bar D^{(*)}(\pi)$~\cite{Shen:2016tzq,Lin:2019qiv,Du:2021fmf}  than into $J/\psi N$.
Second, the production of open-charm final states in high-energy collisions (which has been ignored in the above rough estimates) should have a higher rate than that of a charmonium plus light hadrons because the latter requires the charm and anticharm quarks to be constrained in a smaller phase space.\footnote{Note that there can also be a coupled-channel mechanism in addition to the one shown in Fig.~\ref{fig:diagrams}~(a): open-charm baryon-meson pairs ($\Lambda_c \bar D^{(*)}$, $\Sigma_c^{(*)} \bar D^{(*)}$, etc.) are produced associated with the anti-proton in the final state (or the charge conjugated processes), and then the $J/\psi p$ are produced through rescattering of the open-charm baryon-meson pairs~\cite{Du:2020bqj}. 
Considering such a mechanism would further increase the cross section. Thus, the estimate given in Eq.~\eqref{eq:prediction} should be considered as a rough lower bound.} 
Third, in the most popular hadronic molecular model of the $P_c$ states, they are interpreted as $\Sigma_c^{(*)}\bar D^{(*)}$ hadronic molecules which were first predicted in Ref.~\cite{Wu:2010jy} in the correct energy range, and thus couple strongly to the $\Sigma_c^{(*)}\bar D^{(*)}$; for a determination of the couplings as residues of the scattering amplitudes from the LHCb data~\cite{LHCb:2019kea}, see Ref.~\cite{Du:2021fmf}.

Because the gluons are light flavor isospin and SU(3) singlet, the production rate for the $J/\psi n\bar n$ should be the same as that of the $\jppb$ up to isospin breaking, and the production rates of processes with the nucleons replaced by hyperons, $J/\psi \Lambda \bar \Lambda$, $J/\psi \Xi\bar \Xi$ and $J/\psi \Sigma \bar \Sigma$, should also be of the same order of magnitude for $\sqrt{s}$ sufficiently higher than their thresholds.

To summarize, we present a simple order-of-magnitude estimate of the production cross section of $\ee\to \jppb$, which is expected to be of $\mathcal{O}(4~\text{fb})$ for $e^+e^-$ energies between about 6 and 7~GeV. The cross sections for the production of $J/\psi$ and an isoscalar pair of hyperon and antihyperons should be of the same order. 
The estimates show that the future high-luminosity STCF has a good opportunity to search for the hidden-charm $P_c$ and $P_{cs}$ pentaquarks which were predicted to have a rich spectrum (see, e.g.,~\cite{Dong:2021juy} for a survey of hidden-charm molecular states). 

\begin{acknowledgments}
We are grateful to Kuang-Ta Chao, Hai-Ping Peng, Yu-Jie Zhang and Xiao-Rong Zhou for helpful discussions. In particular, we thank Yu-Jie Zhang for providing his NRQCD result. This project is supported in part by the National Key R\&D Program of China under Grant No. 2023YFA1606703; by the National Natural Science Foundation of China (NSFC) under Grants No. 12125507, No. 12447101, and No. 12361141819; and by the Chinese Academy of Sciences under Grant No. YSBR-101.
\end{acknowledgments}

\appendix
\section{LO calculation of $\sigma(e^+e^- \to J/\psi gg)$}\label{app:A}

\begin{figure}[tb]
    \centering
    \includegraphics[width=0.6\textwidth]{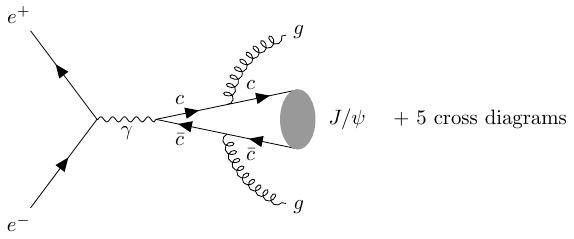}
    \caption{Feynman diagrams for the $J/\psi gg$ production in $e^-e^+$ collisions. }
    \label{fig:feynman}
\end{figure}

Here we present the LO NRQCD calculation of $\sigma(e^+e^- \to J/\psi gg)$. 
The LO Feynman diagrams for this process are shown in Fig.~\ref{fig:feynman}.
This cross section has been calculated in Ref.~\cite{Keung:1980ev}, and it can be expressed as~\cite{Yuan:1997sn}
\begin{equation}
    \frac{d\sigma(e^+e^- \to J/\psi gg)}{\sigma_{\mu\mu} dz dx_1} = \frac{64 e_c^2 \alpha_s^2}{27} \frac{\langle \mathcal{O}_1^\psi(^3S_1)\rangle}{m^3} r^2 f(z, x_1; r),
\end{equation}
where $\sigma_{\mu \mu}$ is the cross section of the QED process $e^{+} e^{-} \rightarrow \mu^{+} \mu^{-}$, and the function $f(z, x_1; r)$ is given by
\begin{align}
    f(z, x_1; r)= \frac{(2 + x_2)x_2}{(2 - z)^2(1 - x_1 - r)^2} + \frac{(2 + x_1)x_1}{(2 - z)^2(1 - x_2 - r)^2} 
    + \frac{(z - r)^2 - 1}{(1 - x_2 - r)^2(1 - x_1 - r)^2} \notag\\
    + \frac{1}{(2 - z)^2} \left( \frac{6(1 + r - z)^2}{(1 - x_2 - r)^2(1 - x_1 - r)^2}  
    + \frac{2(1 - z)(1 - r)}{(1 - x_2 - r)(1 - x_1 - r)r} + \frac{1}{r}\right),
\end{align}
with the variables defined as follows:
\begin{equation}
    z = \frac{2p \cdot k}{s}, \quad x_i = \frac{2p_i \cdot k}{s}, \quad r = \frac{m^2}{s},\quad m = 2m_c,
\end{equation}
where $k$, $p$, $p_i$ are the four-momenta of the virtual photon, $J/\psi$, and the gluons, respectively.   
The integration region is $2\sqrt{r}\leqslant z\leqslant 1+r  $ and $(2-z-\sqrt{z^2-4r})/2\leqslant x_1\leqslant (2-z+\sqrt{z^2-4r})/2$. 
The color-singlet matrix element $\langle \mathcal{O}_1^\psi(^3S_1)\rangle$ is related to the radial wave function of $J/\psi$ at the origin, and can be extracted from the $J/\psi$ leptonic decay width or lattice calculations~\cite{Bodwin:1996tg,Bodwin:2007fz,Ma:2008gq}. 
Since we are considering energies above 6~GeV, the cross section of di-electron to di-muon can be approximated in the ultra-relativistic limit as
\begin{equation}
    \sigma_{\mu\mu} = \frac{4\pi\alpha_{EM}}{3s}.
\end{equation}

\bibliography{refs.bib}

\end{document}